\documentclass[twocolumn,longbibliography]{revtex4-1}
\pdfoutput=1
\usepackage{amsmath}
\usepackage{amssymb}
\usepackage{graphicx}
\usepackage{xcolor}
\usepackage{verbatim}
\usepackage{marginnote}
\usepackage{mathtools}
\usepackage{tabularx}

\newcommand\bfrho{\boldsymbol{\rho}}

\begin{document}

\title{Sliced Basis Density Matrix Renormalization Group for Electronic Structure}

\author{E.\ Miles Stoudenmire}
\affiliation{Department of Physics and Astronomy, University of California, Irvine, CA 92697-4575 USA}

\author{Steven R.\ White}
\affiliation{Department of Physics and Astronomy, University of California, Irvine, CA 92697-4575 USA}

\date{\today}

\begin{abstract}
We introduce a hybrid approach to applying the density matrix renormalization group (DMRG)
to continuous systems, combining a grid approximation along one direction with
a finite Gaussian basis set along the remaining two directions. 
This approach is especially useful for chain-like molecules, 
where the grid is used in the long direction, and we demonstrate the approach
with results for hydrogen chains. The computational time for this system scales approximately
{\it linearly} with the length of the chain, as we demonstrate with minimal basis set
calculations with up to 1000 atoms, which are near-exact within the basis. 
The linear scaling comes from the combination of localization of the basis and a 
compression method with controlled accuracy for the long-ranged Coulomb terms in the Hamiltonian.
\end{abstract}

\maketitle

In the last decade the density matrix renormalization group (DMRG) has become a
powerful method for computing the electronic structure of molecules
\cite{Chan:2011}.  The now standard quantum chemistry DMRG approach (QCDMRG)
works with a discrete Hamiltonian defined by an orthogonalized, contracted
Gaussian basis set \cite{White:1999c}. For systems with strong correlation,
problems of inaccuracy and poor convergence plaguing other approaches are
not a serious problem for DMRG. But QCDMRG has major limitations associated
with basis set size and dimensionality. Calculation times grow rapidly with
the number of active basis functions, and the current practical limit is about
100-200 basis functions. In addition, there are fundamental limitations for 
DMRG when the transverse size of the system becomes large, which we do not try 
to address here.

The Hilbert space used in QCDMRG is the same as that of the Hubbard model, equating 
a Hubbard site with a single basis function.  However, the rapid scaling of computation 
time with the number of basis functions in QCDMRG does not occur for a one-dimensional 
Hubbard model, for which the calculation time is approximately linear (when keeping a fixed
number of states in DMRG).
The main reason for the poor scaling of QCDMRG is the complexity of the Hamiltonian in 
the basis, particularly the two-electron terms.
The electron-electron Coulomb interaction terms are defined by two-electron integrals
\begin{align}
V_{ijkl} = \int_{\mathbf{r}_1}\int_{\mathbf{r}_2} \frac{\phi_i(\mathbf{r}_1)\phi_l(\mathbf{r}_1) \phi_j(\mathbf{r}_2)\phi_k(\mathbf{r}_2)}{|\mathbf{r}_1-\mathbf{r}_2|}
\end{align}
where the $\phi_i(\mathbf{r})$ are orthonormal basis functions. If the basis
functions are delocalized, as they are when using molecular orbitals from a Hartree Fock calculation, 
the number of significant $V_{ijkl}$ terms scales as
$N_b^4$, where $N_b$ is the number of basis functions. 
This leads to a computation time for QCDMRG which 
scales as $N_b^4 m^2 + N_b^3  m^3$, where $m$ is the number of
many-body states kept. \footnote{In long molecules, with minimal basis
sets, truncation of the interactions can improve the scaling of QCDMRG to $O(N_b^2)$.}

\begin{figure}[t]
\includegraphics[width=\columnwidth]{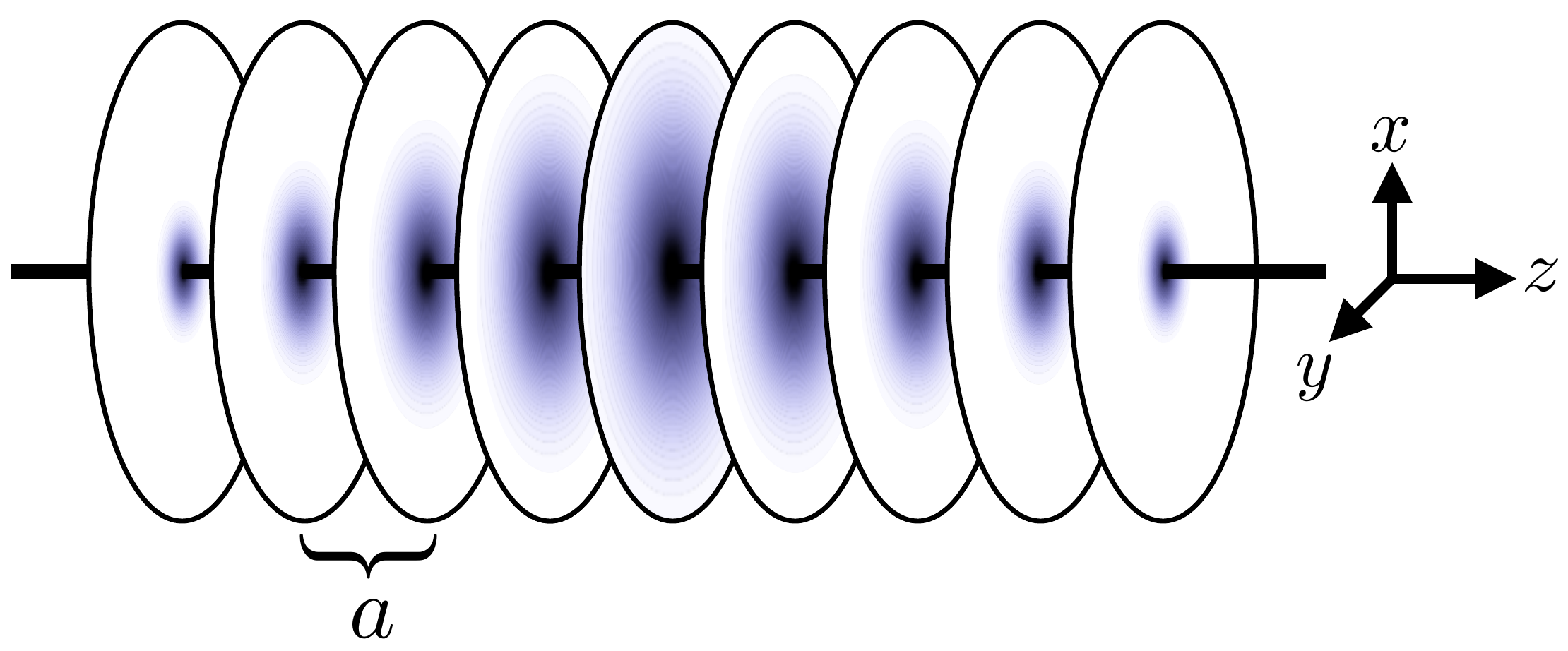}
\caption{The sliced basis set approach can be viewed as finely slicing the continuum into
a collection of parallel two-dimensional planes, each spanned by a small set of transverse functions.}
\label{fig:slices}
\end{figure}

The nonlocality of the orthogonal basis functions also increases the $m$ needed
for a given accuracy. DMRG is a low-entanglement approximation, and the
entanglement of ground states is governed by the \emph{area law}
\cite{Evenbly:2011g,Hastings:2007}. The area law is a property that holds for
ground states  described in terms of local, ``real space'' degrees of freedom.
In a delocalized basis, a volume
law of entanglement holds instead (except for non-interacting systems, a
special point where  the entanglement is zero in the eigenstate basis). To
capture volume-law states, $m$ must grow exponentially with the system size,
even in one dimension.  For this reason, some effort should be made to localize
the basis before applying standard QCDMRG, except on very small molecules. The localization
is always imperfect---the basis functions have oscillating tails which tend to be slowly
decaying.

Hypothetically, one could get rid of both the $N_b^4$ scaling and the increase
in entanglement from extended basis functions by going to a real-space grid
defined by finite differences.  In such a grid the interactions are defined as
$V_{ij} \hat n_i \hat n_j$, where $\hat n_i$ is the density operator on site
$i$.  For model one-dimensional continuum systems, this is currently the most
powerful approach, and we have used it  to simulate systems of 100
pseudo-hydrogen atoms with about 20 grid points per atom
\cite{Stoudenmire:2012d}.  A key part of using a one-dimensional grid is
compressing the interactions by approximating long-range interactions as a sum
of exponentials \cite{Crosswhite:2008,Pirvu:2010}.   With this compression, the
calculation time grows only linearly  with the number of atoms.  The problem
with such a grid approach for three dimensions is that the number of grid
points would be very high, 
for example of order $10^6$ for a system of modest size.

Here we introduce a hybrid approach, which we call sliced basis DMRG (SBDMRG).
Along one particular ``z'' direction we use a grid.  This grid direction is
chosen to be the direction over which the molecule extends furthest. At each
grid point, the remaining transverse dimensions, $x$ and $y$, are captured by a
small number of basis functions derived from standard Gaussian basis sets,
making what we call a ``slice''---see Fig. 1.
The total number of DMRG ``sites'' is
therefore $N_b = N_z N_o$, where $N_z$ is the number of grid points, and $N_o$
is the number of transverse functions (``orbitals'') per grid point. The DMRG
path progresses through all orbitals on a slice,
then moves to the next.  This approach has several major advantages.
First, all interaction terms $V_{ijkl}$ where $i$ and $l$ are not on the same
slice are zero, and similarly for $j$ and $k$. Thus the number of terms scales
as $N_z^2$. Second, the remaining interactions can be compressed very
efficiently, making the dominant part of the calculation time linear in $N_z$.
Third, since there is no spatial extent of the basis functions in the $z$
direction, there is no extra entanglement 
due to nonlocality, potentially reducing the number of states $m$ needed for a given accuracy.

\begin{figure}[t]
\includegraphics[width=\columnwidth]{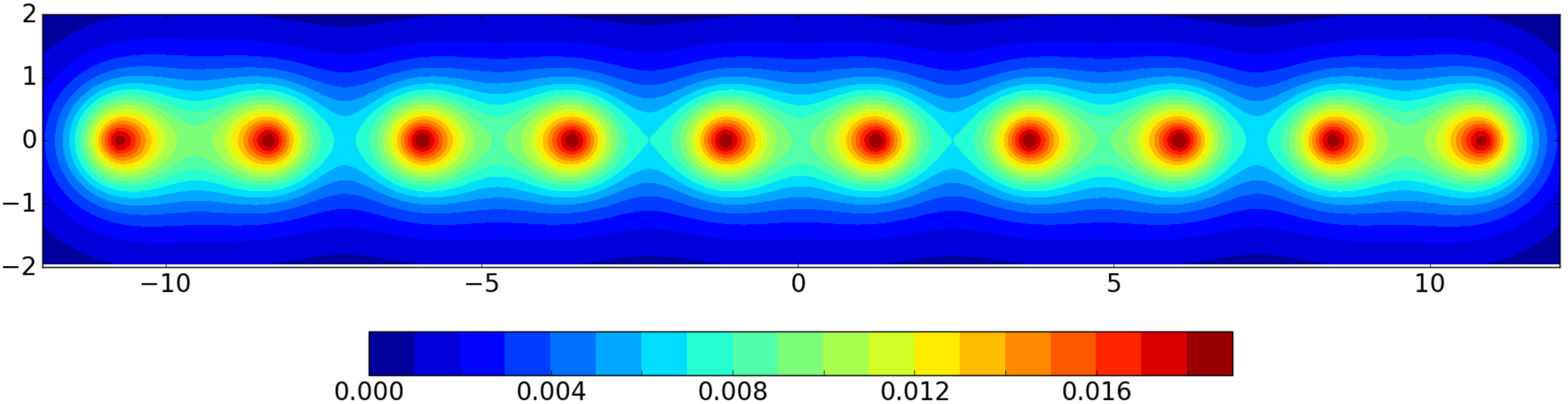}
\caption{Electronic density in the $y-z$ plane of a linear chain of 10 hydrogen atoms, 
    equally spaced at a near neighbor distance $R=2.4$ a.u., calculated in a sliced cc-pVDZ basis (with $N_o=4$). 
A dimerization pattern is visible, induced by
the open ends of the chain, but representing the strong tendency to dimerize into H$_2$ molecules.
}
\label{fig:density}
\end{figure}

We demonstrate our method by simulating linear chains of hydrogen atoms.
Although these are three-dimensional systems, their linear nature makes them
especially well suited for both SBDMRG and QCDMRG. They also exhibit strong
correlation, and can be quite challenging for electronic structure methods.
The electronic density in a plane through the nuclei for a typical calculation
is presented in Fig.~\ref{fig:density}.

To define the sliced basis approach in detail, consider the electronic structure Hamiltonian 
for fixed nuclei in atomic units
\begin{align}
\hat{H}_\text{el} & =  \int_\mathbf{r} \, \hat{\psi}^\dagger_\sigma(\mathbf{r}) \left[ - \frac{1}{2}\nabla^2 + v(\mathbf{r}) \right] \hat{\psi}_\sigma(\mathbf{r}) \nonumber \\
& \mbox{} + \frac{1}{2} \int_{\mathbf{r}, \mathbf{r}^\prime} \frac{1}{|\mathbf{r}-\mathbf{r}^\prime|}  \hat{\psi}^\dagger_\sigma(\mathbf{r}) \hat{\psi}^\dagger_{\sigma^\prime}(\mathbf{r}^\prime) \hat{\psi}_{\sigma^\prime}(\mathbf{r}^\prime)  \hat{\psi}_\sigma(\mathbf{r}) \ . 
\label{eqn:origH}
\end{align}
Summation over spin labels $\sigma$ is implied above and in what follows, and $v(\mathbf{r})$ is the single particle potential generated by the nuclei. 

Along the $z$ direction, we make a grid approximation 
by taking $z_n = n\!\cdot\!a$ with $n$ an integer and $a$ a small grid spacing.
Then on each slice $n$, we introduce a finite, orthonormal basis of functions
$\{ \phi_j(x,y) \}$ where $j=1,2,\ldots,N_o$.
For simplicity, we use the same $N_o$ and functions $\{\phi_j(x,y) \}$ on every slice $n$. 
At a later stage one can perform a change of basis to adapt 
the basis for each slice, possibly reducing the number of functions.
We introduce discrete operators $\hat{c}^\dagger_{n j \sigma}$ and $\hat{c}_{n j \sigma}$
which create and destroy electrons in a slice orbital.
In terms of these operators, the discretized Hamiltonian takes the form
\begin{align}
\hat{H} & =  \frac{1}{2} \sum_{nn^\prime} \sum_{ij} t^{nn^\prime}_{ij} \hat{c}^\dagger_{ni\sigma} \hat{c}_{n^\prime j\sigma} \\
& + \frac{1}{2} \sum_{nn^\prime}\sum_{ijkl} V^{nn^\prime}_{ijkl} \hat{c}^\dagger_{ni\sigma} \hat{c}^\dagger_{n^\prime j\sigma^\prime} \hat{c}_{n^\prime k\sigma^\prime} \hat{c}_{nl\sigma} \ . \label{eqn:HV}
\end{align}
Introducing the notation $\bfrho=(x,y)$ for convenience, the interaction integrals are defined as
\begin{align}
V^{n n^\prime}_{ijkl} & = \int_{\bfrho,\bfrho^\prime} \frac{\phi_i(\bfrho) \phi_j(\bfrho^\prime)\, \phi_{k}(\bfrho^\prime) 
\phi_{l}(\bfrho)}{\sqrt{|\bfrho-\bfrho^\prime|^2 + (z_n - z_{n^\prime})^2}} \ .
\end{align}
Note that the $i,j,k,l$ indices only run over the small number of functions $N_o$ on each slice. Thus,
the Hamiltonian is defined by just $N_z^2 N_o^4$ interaction integrals.
The single-particle couplings are defined to be
\begin{align}
t^{nn^\prime}_{ij} & = 
\delta_{nn^\prime} \int_{\bfrho} \phi_i(\bfrho) \left[ -\frac{1}{2} \nabla^2_{\bfrho} + v(\bfrho,z_n) \right] \phi_j(\bfrho) \\
& - \delta_{ij} \frac{1}{2 a^2} \Delta_{nn^\prime}\ . \label{eqn:gridKE}
\end{align}

Our discrete Hamiltonian treats the $z$-direction kinetic energy terms Eq.~(\ref{eqn:gridKE}) on a 
different footing than the ``integral'' terms.
For the $z$-direction kinetic energy, we treat the basis functions as being smooth functions of $z$, and think of 
the slices as sampling those functions. Thus we use standard finite difference formulas, defined via $\Delta_{nn'}$.
One could take a second order approximation for $\Delta$, with nonzero terms
$\Delta_{nn}=-2$ and $\Delta_{n,n+1} = \Delta_{n+1,n}=1$.
However, to reduce the grid error to $a^4$ we use a fourth-order approximation.  
For the ``integral'' terms, we think of the basis functions as being completely localized and nonoverlapping between
slices, i.e. $\varphi_{nj}(\mathbf{r})=\delta^{\frac{1}{2}}(z-z_n) \phi_j(x,y)$.
This corresponds to taking
\begin{align}
\hat{c}_{n j \sigma} = \sqrt{a} \int_{x,y} \phi_j(x,y)\, \hat{\psi}_{n \sigma}(x,y,z_n) \ .
\end{align}
and then transforming Eq. (\ref{eqn:origH}) accordingly.
The distinct treatments of the terms means that the results are not strictly variational at finite $a$.
However, we find \mbox{finite-$a$} errors for hydrogen chains of only about 0.1 mH per atom for $a=0.1$,
and in the limit of $a \to 0$, the results are variational.

In what follows, we construct the transverse basis functions on a slice 
$\{\phi_j(x,y) \}$ out of standard atom-centered Gaussian basis sets.  
We assume all the atoms are identical.
In going from the spherical symmetry used in standard Gaussians to slices,
we switch to cylindrical symmetry.  Thus, an $S$-function becomes a $\sigma$ function,
$P$-functions become $\pi$ functions, etc.  Whereas there are $2\ell+1$ functions in a spherical set 
with angular momentum $\ell$, there are only two cylindrical functions for any $\ell>0$.
For example, a set of $D$  functions, with coefficient $\zeta$, becomes the two
slice basis functions
\begin{align}
(x^2-y^2) \exp[-\zeta (x^2 + y^2)] \\
2xy \exp[-\zeta (x^2 + y^2)] \ .
\end{align}
We leave out functions like $P_z$, which looks like a $\sigma$ function on a slice, or any
other function looking like a function of smaller $\ell$.
(In principle, $P_z$ could be kept as an additional $S$ function.)
The slice basis functions are only orthogonal between different slices.
This means the functions within each slice must be orthogonalized.

In the parent 3D Gaussian bases, usually some of the functions (particularly $S$-type) are contracted, meaning
out of $N_g$ original Gaussians,
one uses a smaller number $N_o$ of linear combination of functions for each atom:
\mbox{$\phi^j = \sum_{m=1}^{N_g} c^j_m \exp[-\zeta_m (\vec r - \vec r_A)^2]$}
where $j=1\ldots N_o$, and $N_o < N_g$. In this case, to define the transverse
basis on a slice, we follow an approach that is useful very generally: we form
a local orbital density matrix for each slice. 
Let $i$ and $i'$ run over an orthonormal uncontracted basis for the slice at $z_n$, defined by functions $\xi_i(x,y)$.
Let $\phi^k(x,y,z)$ be a particular 3D contracted basis function attached to one of the atoms,
and let 
\begin{equation}
    \eta^k_i = \int_{x,y} \phi^k(x,y,z_n) \xi_i(x,y)
\end{equation}
Then  let
\begin{equation}
    \rho_{ii'} = \sum_k \eta^k_i \eta^k_{i'} .
\end{equation}
The leading eigenvectors of $\rho$ form optimal local functions for representing the contracted 3D basis.
More generally, $\rho$ could come from the interacting ground state, as a block of the single particle reduced
density matrix $\langle c^\dagger_i c_{i'} \rangle$, and we would call the eigenvectors of 
$\rho$ ``slice natural orbitals'' (SNOs).
A subset with only $N_j$ of these 
SNOs would be an ideal reduced local basis. Our procedure for contractions is conceptually similar
to this, but with equal weighting for all 3D contracted basis functions. In this case, for example, the
sharp Gaussians used to represent the nuclear cusps only appear significantly
in the slices close to nuclei.
In our hydrogen chain calculations, if the basis has $N_S$ contracted $S$ functions per atom, 
we keep $N_S$ contracted functions per slice.

We perform DMRG with the Hamiltonian represented as a sum of 
matrix product operators (MPOs), one of which represents
the long-ranged two-electron interactions. For this MPO we use 
a compression technique giving an MPO with matrix dimension $D$ which is nearly
independent of system length, leading to a linear scaling
of the computation time. (The other MPOs, say for $v(\mathbf{r})$, are naturally of constant dimension.)
Consider the
simplest case of a single basis function per slice such that the interaction
part of the Hamiltonian Eq.~(\ref{eqn:HV}) simplifies to
\begin{align}
\sum_{n\leq n'} V_{n n^\prime} \hat{n}_n \hat{n}_{n^\prime} \ . 
\label{eqn:nnV}
\end{align}
Here we will focus on the compression of the upper triangle of the matrix $V_{n n^\prime}$, giving just
an outline; more details are given in Appendix~\ref{app:compression}.
Note that since the local basis varies from slice to slice, $V$ is not translationally
invariant; if it was, an MPO could be constructed based on fitting $V(n-n^\prime)$
to a sum of exponentials \cite{Crosswhite:2008}. We use a more 
general method based on a sequence of
singular value decompositions (SVDs). This is a simplification of more general SVD approaches
for potentially more complicated Hamiltonians \cite{Zaletel:2015i,Chan:2016}.

For a particular diagonal index $k$, let $V^{(k)}$ be the rectangular block of $V$ with
the lower left corner at $V_{kk}$, and extending to the upper right corner of $V$.
An SVD gives
\begin{align}
V^{(k)} = U^{(k)} S^{(k)} W^{(k)}
\label{eqn:Udef}
\end{align}
where $S^{(k)}$ is the diagonal matrix of singular values. The smoothness of $V(n-n^\prime)$ away from the diagonal
makes this SVD have a small number $D$ of significant singular values, allowing us to approximate
$S^{(k)}$ as a $D\times D$ matrix, with appropriate reductions in the number of columns of $U^{(k)}$ and
rows of $W^{(k)}$.

\begin{figure}[b]
\includegraphics[width=\columnwidth]{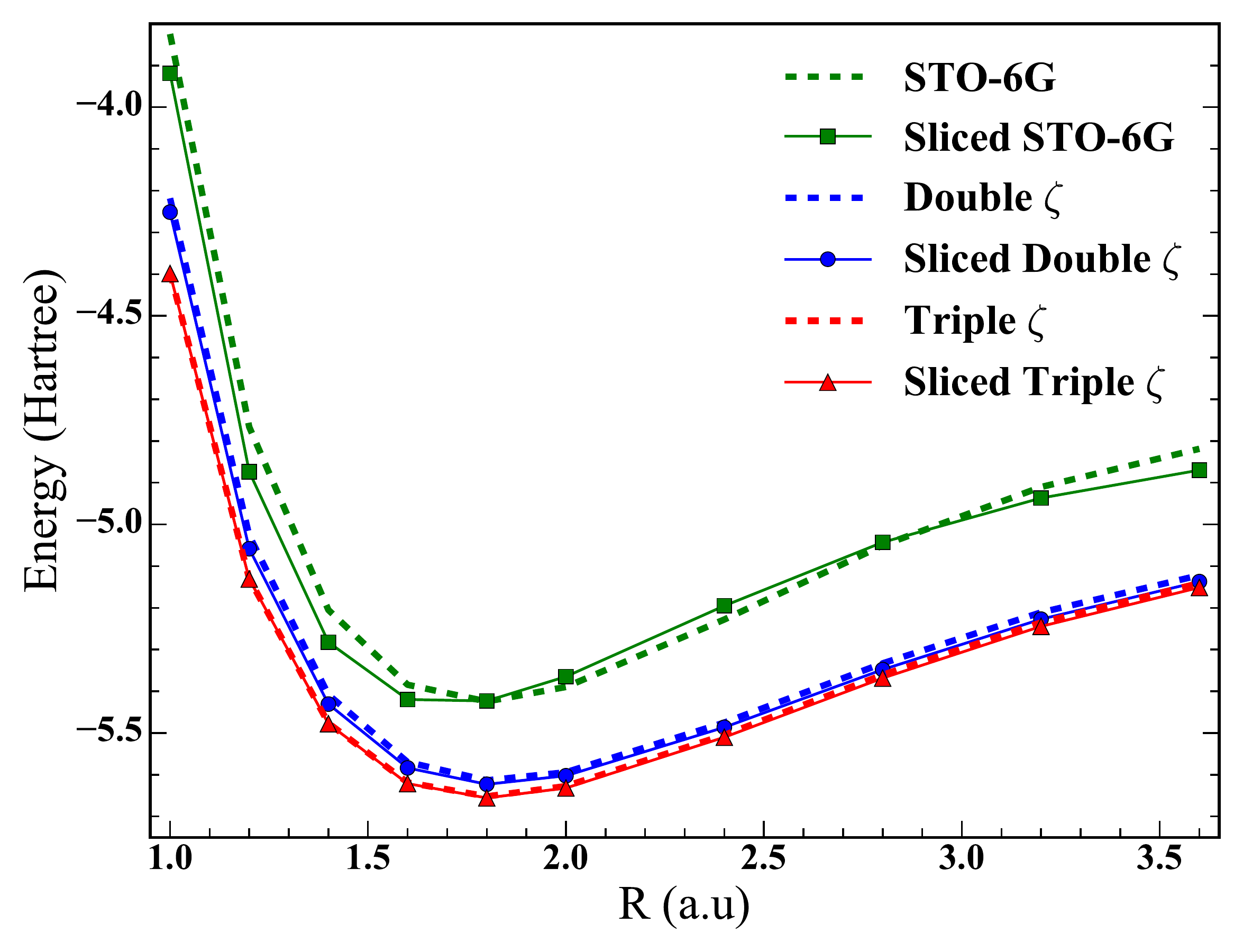}
\caption{Energy of linear chains of 10 hydrogen atoms, equally spaced by a distance
$R$. Dashed lines show results using QCDMRG in standard basis sets. Solid lines with
symbols are SBDMRG results in a sliced version of each basis set.
}
\label{fig:energy}
\end{figure}

This factorized representation at index $k$ can be related to a similar representation at $k+1$.
Define $P(U^{(k)})$ to be the direct sum of $U^{(k)}$ and a $1\times1$
identity matrix, that is add an extra column and row of zeros to the bottom and right of $U^{(k)}$ 
and set the new diagonal element to 1. Then a matrix $X^{(k+1)}$ can be computed such that
\begin{align}
U^{(k+1)} = P(U^{(k)}) X^{(k+1)} \ .
\label{eqn:Ukone}
\end{align}
The matrix $X^{(k+1)}$ is of dimension $(D+1)\times D$.  We see that we can recover all the
$U^{(k)}$ if we know all the $X^{(k)}$ and $U^{(1)}$. Similarly, all of the $W^{(k)}$ 
can be generated in terms of a reverse recursion involving $D \times (D+1)$ matrices $Y^{(k)}$.  
This means we can reconstruct every $V^{(k)}$, and thus the entire $N_z \times N_z$ matrix
$V$ out of the $O(N_z D^2)$ parameters in $X^{(k)}$, $S^{(k)}$, and $Y^{(k)}$.
In Appendix~\ref{app:compression}, we detail how to compute the $X^{(k)}$ and $Y^{(k)}$ matrices,
and show how they lead to an MPO representation of the interactions with MPO matrix dimension
$D+2$.

\begin{figure}[t]
\includegraphics[width=1.1\columnwidth]{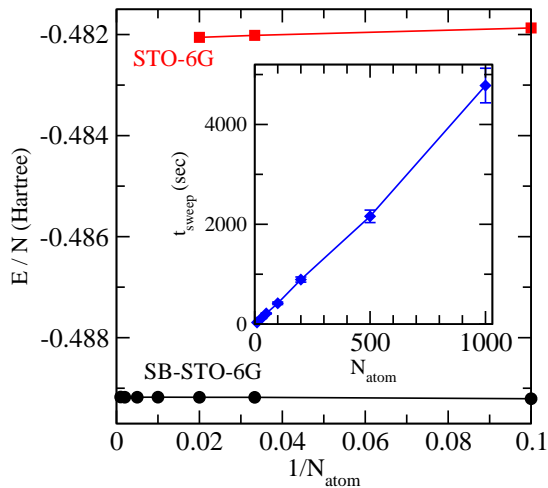}
\caption{Energy per atom of $N$ hydrogen atom chains, equally spaced by a distance
$R=3.6$ a.u.\ using standard versus sliced STO-6G basis sets.
Inset: average time per DMRG sweep with $m=100$, demonstrating linear
scaling up to $N=1000$ atoms.}
\label{fig:scaling}
\end{figure}

In Fig. 3 we show results for chains of 10 equally-spaced hydrogen atoms as a 
function of separation $R$, for several different basis sets with $a=0.1$ and for comparison,
standard QCDMRG results for parent 3D basis sets \cite{Zheng_Comm}.
The STO-6G basis is a 
minimal basis, contracting 6 Gaussians to one function per atom; the sliced version
also has one function per slice.  One can see that the completeness of 
the standard and sliced bases are similar; which basis gives a lower energy varies with $R$.
The double $\zeta$ basis (cc-pVDZ) has five functions per atom \cite{Dunning:1989_I}, and the sliced version has
four per slice (no $P_z$). Here the energies are even closer, but the sliced version is 
consistently slightly lower. The triple $\zeta$ basis (cc-pVTZ) has 14 functions per atom,
or 140 functions total, making this a somewhat challenging QCDMRG calculation.
The sliced version has 9 functions per slice, with up to 561 slices.  
To get the SBDMRG total energy errors to within 1 mH took from 4-10 days (depending on $R$),
with bond dimensions $m \sim 300-1000$,
running on a 2013 quad core Mac mini with 16Gb.
For triple $\zeta$ the sliced and non-sliced energies are also
very close, but with the sliced version slightly lower. All DMRG calculations were 
performed using the ITensor library \cite{ITensor}.

In Fig. 4, we present results for very long chains, demonstrating the linear scaling
of SBDMRG.  These calculations were at the stretched distance $R=3.6$,
using a sliced STO-6G basis with one basis function per slice, and grid spacing $a=0.2$. 
The inset shows the calculation time per sweep on a single core of a 2013 3.5GHz Mac Pro, 
for a sweep
keeping $m=100$ states. The calculation time not only grows very close to linearly in
the number of atoms, it is also quite modest.
The largest system, with 1000 atoms, had over 18,000 sliced basis functions, and an
$m=100$ sweep took a little more than an hour. The number of states kept was slowly ramped
up, with 30 smaller-$m$, faster sweeps occuring before three $m=100$ sweeps.  
Subsequent sweeps up to $m=400$ showed that at $m=100$, the energy per atom was in error
by only 0.06 mH (DMRG error only, excluding the finite basis and finite $a$ errors).  
The main part of the figure shows the energy per site, in comparison with QCDMRG STO-6G. 
The energy results show the modest difference in completeness of STO-6G and sliced
STO-6G, and also demonstrate that the sliced DMRG is converged to high accuracy.

The sliced basis set approach we have introduced here can be seen to be very well suited to
DMRG calculations.  Coupled with a compression method for the interactions, this approach gives
linear scaling of computation time with the length of the system, allowing very long systems
to be treated. This formulation brings DMRG for electronic structure closer to DMRG for models,
and new approaches introduced for models (such as working directly with an infinite chain)
can probably be adapted to SBDMRG with little difficulty. We also anticipate that extending
SBDMRG to more complicated molecules will be reasonably straightforward.

We acknowledge support from the Simons Foundation through the Many-Electron Collaboration,
and from the U.S. Department of
Energy, Office of Science, Basic Energy Sciences under
award \#DE-SC008696.

\bibliography{sbdmrg}

\appendix

\section{Interaction Integrals for Sliced Basis Sets}

Recall that to construct a Hamiltonian in a sliced basis set, one must compute the integrals
\begin{align}
V^{n n^\prime}_{ijkl} & = \int_{\bfrho,\bfrho^\prime} \frac{\phi_i(\bfrho) \phi_j(\bfrho^\prime)\, \phi_{k}(\bfrho^\prime) 
\phi_{l}(\bfrho)}{\sqrt{|\bfrho-\bfrho^\prime|^2 + (z_n - z_{n^\prime})^2}} \label{eqn:Vint}
\end{align}
for the interaction terms and 
\begin{align}
\tilde{t}^{nn}_{ij}  & =  \int_{\bfrho} \phi_i(\bfrho) \left[ -\frac{1}{2} \nabla^2_{\bfrho} \right] \phi_j(\bfrho) \label{eqn:tintKE} \\
& \mbox{} + \int_{\bfrho} \phi_i(\bfrho) \left[ v(\bfrho,z_n)  \right] \phi_j(\bfrho) \label{eqn:tintV}
\end{align}
for the single-particle terms. (Recall the full expression for $t^{nn}_{ij}$ includes the grid kinetic energy \mbox{$t^{nn}_{ij}=\tilde{t}^{nn}_{ij} - \frac{1}{2 a^2} \Delta_{nn}$}.)

To use a sliced basis, we need to evaluate integrals between basis function representing: 
\begin{enumerate}
\item the overlap of two nonorthogonal function on a slice 
\item kinetic energy matrix elements on a slice, Eq.~(\ref{eqn:tintKE})
\item single particle potential matrix elements from the Coulomb potential of the nuclei, Eq.~(\ref{eqn:tintV})
\item the two particle terms $V^{nn^\prime}_{ikjl}$, Eq.~(\ref{eqn:Vint})
\end{enumerate}
The integrals for (1) and (2) for Gaussian functions have simple analytic
formulas.  The matrix elements (3) can be considered a limiting case of (4), 
where we consider a nucleus as an $S$-type Gaussian of 
vanishing width on one slice, and then the terms from the second coordinate define the one particle potential;
thus we need only consider case (4).

\subsection{Gaussian Fitting For $\ell > 0$ Integrals}

For $S$ functions, the $V^{nn^\prime}_{ijkl}$ have analytic formulas.  However, for other types of orbitals, the formulas get 
both tedious to derive and very time consuming to evaluate. Instead, we implemented another approach: fit the 
function $1/r$ to a sum of Gaussians
\begin{equation}
\frac{1}{r} \approx \sum_{i=1}^{P} c_i \exp(-a_i r^2) \ .
\end{equation}
The widths of these Gaussians $a_i^{-1/2}$ were taken to be equally spaced on a logarithmic scale, except for
the ten largest widths, which were optimized over both $a_i$ and $c_i$. 
Using $P=220$, we obtained a fit good to $O(10^{-10})$ over the range $10^{-8}$ to $10000$.  
The integrals were evaluate by taking the sum over $i$ outside the integrals, turning them into simple analytic
Gaussian integrals which also separated by dimension $x, y, z$.  The separation meant that the integral
formulas for each single dimension could be calculated and stored quickly, and then each $V^{nn^\prime}_{ijkl}$
evaluation could be done as a loop of length $P$ involving only multiplications and additions, making it very
fast.

\subsection{Smoothing Procedure for Integrals with Cusps}

\begin{figure}[t]
\includegraphics[width=0.9\columnwidth]{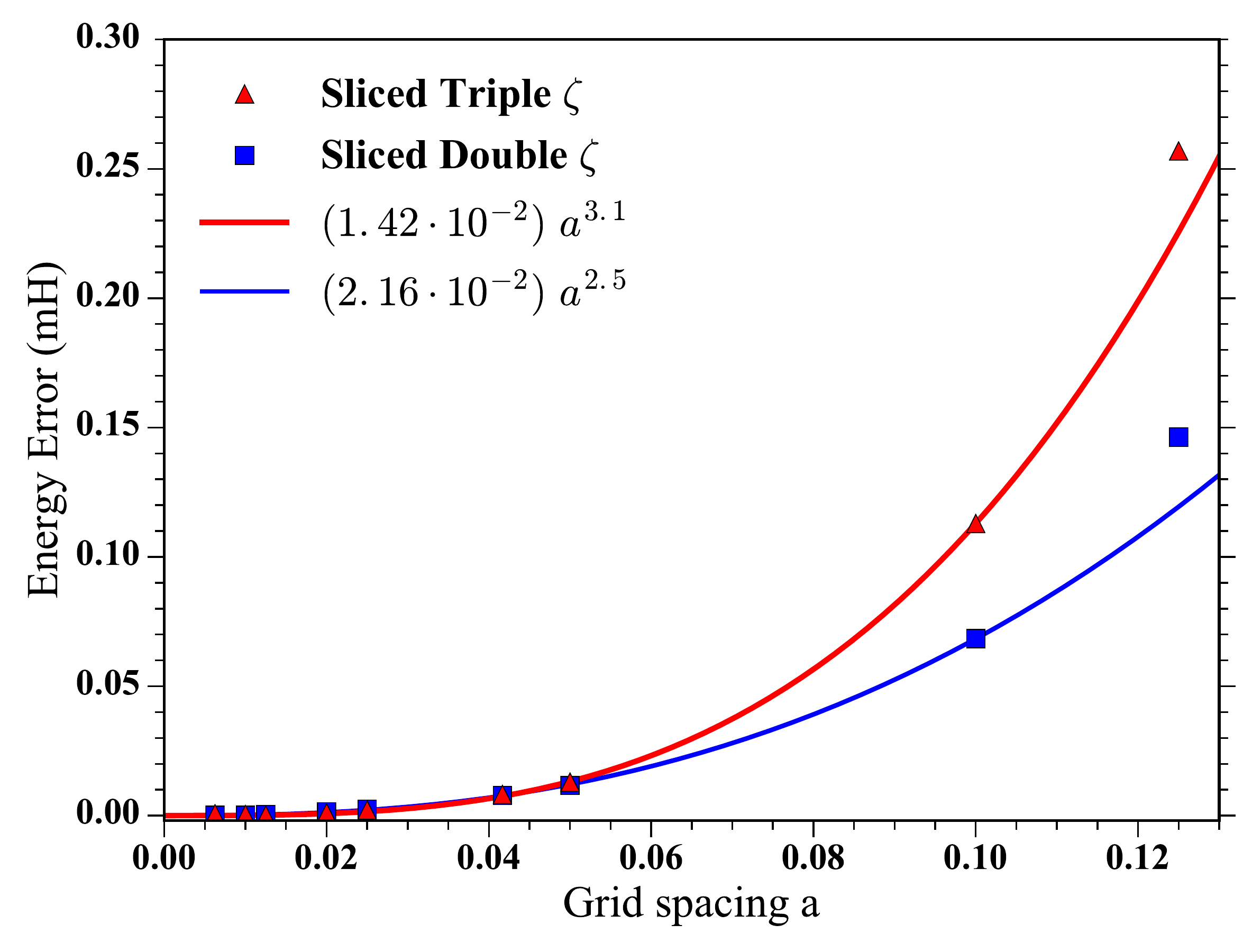}
\caption{Scaling of the energy error of single hydrogen atom energies as a function
of grid spacing $a$. Data shown are for the basis sets sliced double $\zeta$ and sliced triple $\zeta$. 
For $a=0.2$ (not shown), the energy error rises to about 1 mH.}
\label{fig:scaling}
\end{figure}

If a continuous function does not have any frequency components above $\pi/a$,
sampling it with grid spacing $a$ is exact.  
In contrast, sampling a function with a slope discontinuity
leads to errors in the function of order $a$. The divergence of the $1/r$ interaction at short distances
makes some of the two electron interaction integrals have slope discontinuities at $z = z'$.

To accelerate the convergence with $a$, we adopt a pre-filtering technique, which is done before any
contractions, when the integrals are still a function of $z-z'$.  The interaction is first computed at
a finer grid spacing of $2^{-r} a$ for a small integer $r$. Then the interactions are put through a low-pass
filter and factor-of-two decimation $r$ separate times, giving a final spacing of $a$.  The low pass filter
is designed to reproduce exactly all frequencies up to half the maximum frequency.  Thus this smoothing
procedure does not alter any low frequency parts of the interaction, but smoothly removes components at frequencies
higher than $\pi/a$.  The same smoothing procedure is also used for the
nucleus-electron interaction integrals.  We tested the accuracy of this
procedure on H$_2$, and found $r=3$ nicely accelerates convergence with $a$
while not increasing the computation time too much. The errors in the resulting
total energies shown in Fig.~(\ref{fig:scaling}) scale approximately as $a^{2.5}$ to $a^{3.1}$, and are approximately 0.1 mH per atom at $a=0.1$.

\section{SVD Compression of Long-Range Interactions \label{app:compression}}

In this section we give a more detailed discussion of the compression algorithm
for long-range interactions described in the main body of the paper. The simplest
case is compressing the interaction part of the sliced basis set Hamiltonian for the 
case of one transverse function per slice.
\begin{align}
\sum_{n \leq n^\prime} V_{nn^\prime} \hat{n}_{n} \hat{n}_{n^\prime}
\end{align}
Later below we discuss how to generalize the compression for the case of multiple
transverse functions. 

The basic idea of the compression algorithm is to use the singular value decomposition (SVD)
to compress each of the rectangular blocks $V^{(k)}$ of the matrix $V$ extending from the 
element $V_{kk}$ to the upper-right corner of $V$. 
As a motivation, consider the case where the interactions decay exponentially:
\begin{align}
V_{nn^\prime} = \lambda^{|n-n^\prime|} \ .
\end{align}
Restricting $V$ to an upper-right block constrains $n^\prime \geq n$, in which case $V$ factorizes as
\begin{align}
V_{nn^\prime} = \lambda^{|n-n^\prime|} = \lambda^{n^\prime-n} = \lambda^{-n} \lambda^{n^\prime}\ \ \ (n^\prime \geq n)\ .
\end{align}
This factorization into the outer product of two vectors implies that each upper-right block $V^{(k)}$ has only one non-zero 
singular value (is rank 1) and will be maximally compressed by an SVD. The interaction matrix
for a real system will be more complicated, but if one can approximate it as a sum of exponentials,
then the number of significant singular values of $V^{(k)}$ should remain small. In practice, the 
SVD can uncover better compression strategies than just a sum of real exponentials.

\subsection{Algorithm for an $N\times N$ Matrix \label{sec:algorithm}}

\begin{figure}[b]
\includegraphics[width=0.7\columnwidth]{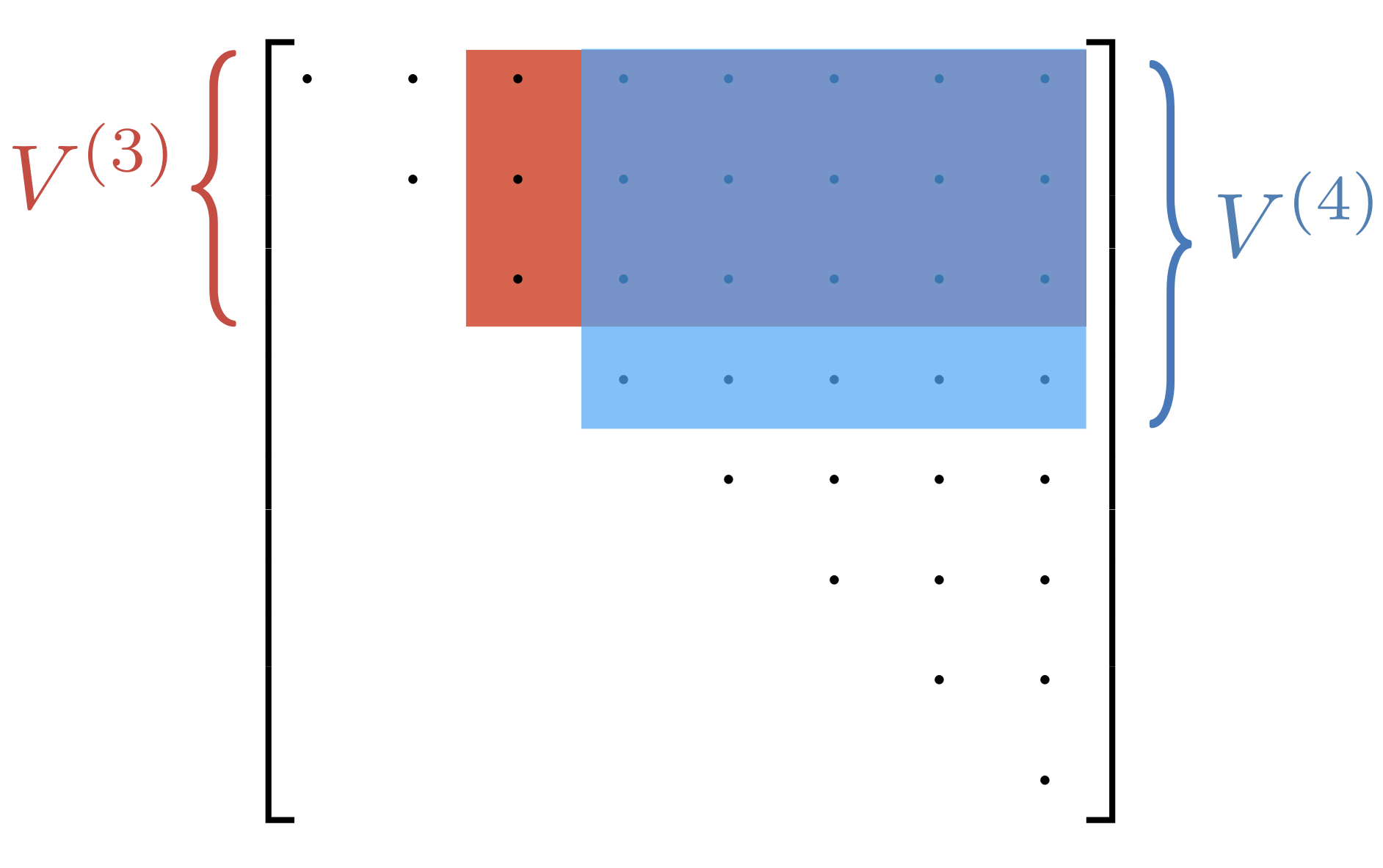}
\caption{Blocks $V^{(3)}$ and $V^{(4)}$ of an $8\times 8$ upper-triangular matrix $V$.}
\label{fig:blocks}
\end{figure}

First we will detail the compression algorithm for the case where $V_{n n^\prime}$ is just an 
$N\times N$ matrix, and later generalize to the case where $V$ is a tensor (the latter 
corresponding in SBDMRG to having multiple functions on each slice).
The compression deals with the upper-right blocks $V^{(p)}$ of $V$, defined such that
\begin{align}
V^{(p)}_{r,c} = V_{r,(c+p-1)}
\end{align}
where $r=1,2,...,p$ and $c=1,2,\ldots,(N-p+1)$, see Fig.~\ref{fig:blocks}.

For each of these blocks we define the matrices $U^{(p)}$, $S^{(p)}$, and $W^{(p)}$ by an
SVD of $V^{(p)}$:
\begin{align}
V^{(p)} = U^{(p)} S^{(p)} W^{(p)} \ . \label{eqn:blockSVD}
\end{align}
The matrix $S^{(p)}$ is diagonal and contains the singular values. Assuming the smoothness
of $V_{nn^\prime}$ away from the diagonal makes the $V^{(p)}$ have only $D$ significant
singular values, the compression is achieved by truncating $S^{(p)}$ to be only a $D\times D$
matrix, reducing the corresponding columns of $U^{(p)}$ and rows of $W^{(p)}$.

We next seek a way to relate the SVD of any one of the blocks $V^{(p)}$ to another 
block $V^{(p+1)}$. It is helpful to define the following additional notation:
\begin{enumerate}
\item Define $C^-(M)$ to be the matrix $M$ with the first column removed (making a smaller matrix).
\item Define $M \oplus r$ to be $M$ with an extra row $r$ added at the bottom ($r$ is a vector).
\item For an $n\times m$ matrix $M$, define $P(M)$ to be $M$ with an extra row and column added at the 
bottom and right. The extra matrix elements are zero, except for a 1 on the diagonal (at position $(n+1),(m+1)$).
\item Define $r^{(p)}$ to be the bottom row of $V^{(p)}$.
\end{enumerate}
Then it follows that
\begin{align}
    V^{(p+1)} &= C^-(V^{(p)}) \oplus r^{(p+1)} \label{eqn:addrow}\\
     &= \big[U^{(p)} S^{(p)} C^-(W^{(p)})\big] \oplus r^{(p+1)}\\
     &= P(U^{(p)}) \big[S^{(p)} C^-(W^{(p)}) \oplus r^{(p+1)} \big] \label{eqn:preX} .
\end{align}
Writing the SVD of the matrix in square brackets in Eq.~(\ref{eqn:preX}) as $X^{(p+1)} S^{(p+1)} W^{(p+1)}$, we find that
we have obtained the SVD of $V^{(p+1)}$, with 
\begin{align}
    U^{(p+1)} = P(U^{(p)}) X^{(p+1)}
\label{eqn:Upone}
\end{align}
Each matrix $X^{(p)}$ is of dimension $(D+1) \times D$.  We see that we can recover all the
$U^{(p)}$ if we know all the $X^{(p)}$ plus $U^{(1)}$.  A similar calculation gives all the $W^{(p)}$ in
terms of a reverse recursion involving $D \times (D+1)$ matrices $Y^{(p)}$.  
This means we can reconstruct the entire $N \times N$ matrix
$V(n-n^\prime)$ out of the $O(N D^2)$ parameters in $X^{(p)}$, $S^{(p)}$, and $Y^{(p)}$.

In practice, to obtain the fully compressed representation of $V$, it is useful to start
by computing the SVD of $V^{(2)}$ (the SVD of $V^{(1)}$ is trivial). The initial SVD 
has a cost only linear in $N$ since $V^{(2)}$ is a $2\times N$ matrix. To compute
the $X^{(p)}$, one computes SVDs of the matrices $[S^{(p)} C^-(W^{(p)}) \oplus r^{(p+1)}]$
which are of dimension $(D+1)\times(N-p+1)$. Thus the cost for each of these SVDs scales as
$D^2 N$ (assuming the entries of the matrix $V$ have already been computed). For a 
non-translationally invariant system, one must perform $N$ such SVDs, making the total cost
$D^2 N^2$. But the compression algorithm only has to be performed once, and thus does not dominate 
the scaling of a SBDMRG calculation. To achieve a linear scaling of the compression algorithm,
one could start with a translationally invariant basis such that the SVD Eq.~(\ref{eqn:blockSVD})
is the same for every block of $V$. Following the compression, the basis can be contracted
to a smaller number of functions in a non-translationally-invariant manner on each slice.

\subsection{MPO Form of Compressed Interactions}

A matrix product operator (MPO) is a compact rewriting of a sum of operators as a 
tensor network. An MPO resembles a matrix product state (MPS), but in an MPO each tensor 
has two physical indices. Thus each MPO tensor can be viewed as an operator valued matrix,
which will be the notation we use below.
Representing the Hamiltonian as an MPO, or as a sum of MPOs, not only makes a code
more generic and flexible, but can also make calculations more efficient.

Any sum of finite-range operators can be written exactly as an MPO using
well-known conventions, which results in internal MPO indices whose sizes 
depend linearly on the range of the operators \cite{McCulloch:2007,Crosswhite:2008a}.
However, such an approach fails to be efficient when Hamiltonian 
terms do not have strictly finite support.

An interesting extension of the finite-range MPO construction allows 
MPOs to exactly capture sums of operators whose coefficients decay as pure 
exponentials \cite{McCulloch:2008,Crosswhite:2008}. By fitting other kinds
of long-range terms, such as power-law decaying terms,
to a sum of exponentials \cite{Pirvu:2010}, they can be 
approximated by MPOs in an efficient way.

But the exponential fitting approach leaves much to be desired. The best
quality fits involve complex exponents, yet working with complex numbers incurs
significant computational costs. Using a two-dimensional real matrix representation
of the complex numbers avoids this issue, but complicates the method. 
Setting up the fits and the logic of the exponential decays for ladders
and other quasi-one-dimensional systems with unit cells is also quite difficult.

Here we present an alternate approach to approximating sums of long-range
operators as MPOs based on the SVD based compression algorithm discussed above.
The approach here is closely related to the one proposed in Ref.~\onlinecite{Chan:2016},
especially in terms of the final MPO produced. But the present approach has some extra efficiencies
arising from step that computes each $X^{(p)}$  from a matrix with only $(D+1)$ rows
defined in square brackets Eq.~(\ref{eqn:preX}). The cost of 
each associated SVD is at most linear in $N$,
whereas the proposal in Ref.~\onlinecite{Chan:2016} requires 
SVDs scaling as $N^3$. Both approaches
are also related to a very general proposal for compressing 
MPOs in Ref.~\onlinecite{Zaletel:2015i}.

In this section, we want to use the compression algorithm to produce an MPO for
the sum of operators
\begin{align}
\hat{V} = \sum_{n \leq n^\prime} V_{nn^\prime}\, \hat{n}_n \hat{n}_{n^\prime} \ .
\end{align}
where $n,n^\prime = 1,2,\ldots,N$.
An MPO representation of $\hat{V}$ can be written as
\begin{align}
\hat{V} = \sum_{\{\alpha\}} \hat{M}^{(1)}_{\alpha_1} \hat{M}^{(2)}_{\alpha_1 \alpha_2} \hat{M}^{(3)}_{\alpha_2 \alpha_3} \cdots \hat{M}^{(N)}_{\alpha_{N-1}}
\end{align}
where each $M^{(n)}$ is an operator-valued matrix.

To make the following expressions more compact, it is convenient to define $\Omega^{(p)} = X^{(p)} S^{(p)} W^{(p)}$.
Define the first MPO tensor to be:
\begin{align}
\hat{M}^{(1)} = \begin{bmatrix}
V_{11} (\hat{n}_1)^2 & X^{(1)}_{11} \hat{n}_1 & \hat{I}_1
\end{bmatrix} \ ,
\label{eqn:MPO1_1}
\end{align}
noting that $X^{(1)}\stackrel{\text{def}}{=} U^{(1)} = 1$.
Define the second MPO tensor to be:
\begin{align}
\hat{M}^{(2)} = \begin{bmatrix}
\hat{I}_2 & 0                & 0                      & 0                    \\
\Omega^{(2)}_{11} \hat{n}_2 & X^{(2)}_{11} \hat{n}_2 & X^{(2)}_{12} \hat{n}_2 & 0 \\
V_{22} (\hat{n}_2)^2 & X^{(2)}_{21} \hat{n}_2 & X^{(2)}_{22} \hat{n}_2 & \hat{I}_2 
\end{bmatrix}\ .
\label{eqn:MPO1_2}
\end{align}
And define the third MPO tensor to be:
\begin{align}
\hat{M}^{(3)} = \begin{bmatrix}
\hat{I}_3 & 0                & 0                      & 0                & 0   \\
\Omega^{(3)}_{11} \hat{n}_3 & X^{(3)}_{11} \hat{n}_3 & X^{(3)}_{12} \hat{n}_3 & X^{(3)}_{13} \hat{n}_3 & 0 \\
\Omega^{(3)}_{21} \hat{n}_3 & X^{(3)}_{21} \hat{n}_3 & X^{(3)}_{22} \hat{n}_3 & X^{(3)}_{23} \hat{n}_3 & 0 \\
V_{33} (\hat{n}_3)^2 & X^{(3)}_{31} \hat{n}_3 & X^{(3)}_{32} \hat{n}_3 & X^{(3)}_{33} \hat{n}_3 & \hat{I}_3 
\end{bmatrix}
\label{eqn:MPO1_3}
\end{align}
The general pattern for site $n$ is:
\begin{widetext}
\begin{align}
\hat{M}^{(n)} = \begin{bmatrix}
\hat{I}_n & 0                & 0                      & 0                & 0   & 0 \\
\Omega^{(n)}_{11} \hat{n}_n & X^{(n)}_{11} \hat{n}_n & X^{(n)}_{12} \hat{n}_n & \cdots & X^{(n)}_{1D} \hat{n}_n & 0 \\
\Omega^{(n)}_{21} \hat{n}_n & X^{(n)}_{21} \hat{n}_n & X^{(n)}_{22} \hat{n}_n & \cdots & X^{(n)}_{2D} \hat{n}_n & 0 \\
\vdots                   & \vdots                 & \ddots                 &  & \vdots                       & \vdots \\
\Omega^{(n)}_{D1} \hat{n}_n & \vdots                 &                        &        & \vdots & 0 \\
V_{nn} (\hat{n}_n)^2 & X^{(n)}_{(D+1)1} \hat{n}_n & X^{(n)}_{(D+1)2} \hat{n}_n & \cdots & X^{(n)}_{(D+1)D} \hat{n}_n  & \hat{I}_n 
\end{bmatrix}\ .
\label{eqn:MPO1_gen}
\end{align}
\end{widetext}
From which we see the MPO has a matrix dimension of $(D+2)$.

Expanding this MPO, we can see that it represents $V$ as a sum of terms of the form
\begin{align}
\hat{V} & = \!\!\!\!\!\sum_{n < n^\prime, \{\alpha\}}\!\!\!\!\! \left(X^{(n)}_{(D+1)\alpha_n} \tilde{X}^{(n+1)}_{\alpha_{n} \alpha_{n+1}}
\cdots \tilde{X}^{(n^\prime-1)}_{\alpha_{n'-2}\alpha_{n'-1}} \tilde\Omega^{(n')}_{\alpha_{n'-1}1} \right) \hat{n}_{n} \hat{n}_{n^\prime} \label{eqn:MPOexpand} \\
& + \sum_{n} V_{nn} (\hat{n}_n)^2
\end{align}
where the notation $\tilde{M}$ means the first $D$ rows of a matrix $M$ (either $X$ or $\Omega$).
To see how the expression in Eq.~(\ref{eqn:MPOexpand}) recovers the matrix $V_{nn^\prime}$, note that row $(D+1)$ 
of each matrix $X^{(n)}$ is identical to row $n$ of $U^{(n)}$. Also note that
\begin{align}
\sum_{\alpha_n} U^{(n)}_{r\alpha_n} \tilde{X}^{(n+1)}_{\alpha_{n} \alpha_{n+1}} = U^{(n+1)}_{r\alpha_{n+1}}
\end{align}
for any $r \leq n$; the above equation can be seen to hold by omitting the last row of each of the matrices
in Eq.~(\ref{eqn:Upone}).
It follows that
\begin{align}
& \sum_{\{\alpha\}=1}^D X^{(n)}_{(D+1)\alpha_n} \tilde{X}^{(n+1)}_{\alpha_{n} \alpha_{n+1}} \cdots \tilde{X}^{(n^\prime-1)}_{\alpha_{n^\prime-2}\alpha_{n^\prime-1}} \tilde\Omega^{(n')}_{\alpha_{n'-1}1} =  \\
& = \sum_{\{\alpha\}=1}^D U^{(n)}_{n\alpha_n} \tilde{X}^{(n+1)}_{\alpha_{n} \alpha_{n+1}} \cdots \tilde{X}^{(n^\prime-1)}_{\alpha_{n^\prime-2}\alpha_{n'-1}} \tilde\Omega^{(n^\prime)}_{\alpha_{n'-1}1}  \\
& = \sum_{\{\alpha\}=1}^D U^{(n^\prime-1)}_{n\alpha_{n^\prime-1}} \tilde\Omega^{(n^\prime)}_{\alpha_{n'-1}1}  \\
& = \sum_{\{\alpha\}=1}^D U^{(n^\prime-1)}_{n\alpha_{n^\prime-1}} \tilde{X}^{(n')}_{\alpha_{n'-1}\alpha_{n'}} S^{(n')}_{\alpha_{n'} \alpha^{\prime}_{n'}} W^{(n^\prime)}_{\alpha^\prime_{n^\prime}1}  \\
& = \sum_{\{\alpha\}=1}^D U^{(n^\prime)}_{n\alpha_{n'}} S^{(n')}_{\alpha_{n'} \alpha^{\prime}_{n'}} W^{(n^\prime)}_{\alpha^\prime_{n^\prime}1}  \\
& = V_{nn^\prime}
\end{align}

\subsection{Generalization to Multiple Transverse Functions}

For a sliced basis set with multiple transverse functions $\phi_{j}(x,y)$ on each slice,
the interaction terms have the general form
\begin{align}
\frac{1}{2} \sum_{nn^\prime}\sum_{ijkl} V^{nn^\prime}_{ijkl} c^\dagger_{ni\sigma} c^\dagger_{n^\prime j\sigma^\prime} c_{n^\prime k\sigma^\prime} c_{nl\sigma} \ .
\end{align}
where $n,n^\prime=1,2,\ldots,N_z$ and $i,j,k,l=1,2,\ldots,N_o$.
Thus to compress these interactions one must compress the tensor $V^{nn^\prime}_{ijkl}$.
Because the indices $i,l$ label functions on slice $n$ and $j,k$ functions on slice $n^\prime$,
reshape the tensor to an $(N_z N_o^2)\times(N_z N_o^2)$ matrix 
\begin{align}
V_{(nil)(n^\prime jk)} = V^{nn^\prime}_{ijkl} \ .
\end{align}
Then we can use a similar compression algorithm as that described above, with the key difference that
one defines blocks of $V$ according to the $n,n^\prime$ indices, treating the $i,l$ or $j,k$
indices as a ``unit cell'' for each value of $n$ or $n^\prime$. So in contrast to the previous algorithm,
where one would add a single row of $V$ in Eq.~(\ref{eqn:addrow}), for example, in the more general algorithm
one adds $N_o^2$ rows of $V_{(nil)(n^\prime jk)}$.

The SVD one wants to obtain for each block $V^{(p)}$ of $V$ is of the form
\begin{align}
V^{(p)}_{(ril)(cjk)} & = \sum_{\alpha,\alpha^\prime=1}^{D} U^{(p)}_{(ril) \alpha} S^{(p)}_{\alpha \alpha^\prime} W^{(p)}_{\alpha^\prime (cjk)}
\end{align}
where $r=1,2,...,p$ and $c=1,2,\ldots,(N-p+1)$, and $i,l$ label the functions on slice $r$ while
$j,k$ label the functions on slice $c$.

To compute matrices $X^{(p)}$ relating the SVD at one slice to that at another, make the following
definitions:
\begin{enumerate}
\item Define $C^-_{N^2_o}(M)$ to be the matrix $M$ with the first $N_o^2$ columns removed.
\item Define $A\oplus B$ for an $a\times m$ matrix $A$ and an $b \times m$ matrix $B$ to be
      the $(a+b)\times m$ matrix whose first $a$ rows are those of $A$ and last $b$ rows are those of $B$.
\item Define $P_{N^2_o}$ for an $n\times m$ matrix $M$ to be the direct sum of $M$ and an $N_o^2\times N_o^2$
      identity matrix. That is, append $N^2_o$ rows and columns to $M$ that are zero except for the diagonal
      elements which equal 1.
\item Define $v^{(p)}$ to be the last $N^2_o$ rows of $V^{(p)}$.
\end{enumerate}
Then the block $V^{(p+1)}$ is given by
\begin{align}
V^{(p+1)} & = C^-_{N^2_o}(V^{(p)}) \oplus v^{(p+1)} \\
          & = \big[U^{(p)} S^{(p)} C^-_{N^2_o}(W^{(p)}) \big] \oplus v^{(p+1)} \\
          & = P_{N^2_o}(U^{(p)}) \big[S^{(p)} C^-_{N^2_o}(W^{(p)}) \oplus v^{(p+1)} \big] \ . \label{eqn:preXNo}
\end{align}
By computing an SVD of the matrix in square brackets in Eq.~(\ref{eqn:preXNo}) above, and writing this SVD
as $X^{(p+1)} S^{(p+1)} W^{(p+1)}$, it follows that
\begin{align}
U^{(p+1)} = P(U^{(p)}) X^{(p+1)}
\end{align}
similar to the algorithm for the $N_o=1$ case in Section~\ref{sec:algorithm}. It is helpful to note that the last
$N^2_o$ rows of each matrix $X^{(p)}$ correspond to the last $N^2_o$ rows of $U^{(p)}$, which correspond
to the indices $i,l$ labeling functions on slice $p$.

Finally, for the next section on constructing an MPO, it will be convenient to define
\begin{align}
\Omega^{(p)}_{\alpha_{p-1}(cjk)} = X^{(p)}_{\alpha_{p-1}\alpha'_p} S^{(p)}_{\alpha'_p \alpha_p} W^{(p)}_{\alpha_p (cjk)}
\end{align}
\mbox{} \\

\subsection{MPO For $N_o > 1$ Orbitals on Each Slice}

Consider the case $N_o = 3$ and $D=2$. 
To lighten the notation, we consider a particular slice $p$, suppressing the label $(p)$
and assuming all MPO matrices and matrices $X_{\alpha_{p-1}\alpha_p}$,$U^{il}_{r \alpha_p}$,$\Omega^{jk}_{\alpha_{p-1} c}$ are all 
associated with the same slice $p$.
Subscripts on MPO matrices and on operators indicate the orbital number within the slice $p$.
The entire MPO is formed by repeating these $N_o$ matrices for all $N_z$ slices, together with the boundary
conditions given later below.

The MPO matrix for the first orbital on a slice is:
\begin{widetext}
\setcounter{MaxMatrixCols}{30}
\begin{align}
\hat{M}_1 = \begin{bmatrix}
I_1 & V_{pp} (n_1)^2  & 0    & 0   & n_1 &0&0&0&0&0&c^\dagger_{1\uparrow}&0&0&0&-c_{1\uparrow}&0&\cdots\\
0   & I_1             & 0    & 0   & 0   &0&0&0&0&0&0&0&0&0&0&0&\cdots\\
0   & \Omega^{11}_{11} n_1 & I_1  & 0   & 0   &0&0&0&\Omega^{12}_{11} c^\dagger_{1\uparrow} & \Omega^{13}_{11} c^\dagger_{1\uparrow}&0&0&-\Omega^{21}_{11} c_{1\uparrow} & -\Omega^{31}_{11} c_{1\uparrow}&0&0&\cdots\\
0   & \Omega^{11}_{21} n_1 & 0    & I_1 & 0   &0&0&0&\Omega^{12}_{21} c^\dagger_{1\uparrow} & \Omega^{13}_{21} c^\dagger_{1\uparrow}&0&0&-\Omega^{21}_{21}c_{1\uparrow}&-\Omega^{31}_{21}c_{1\uparrow}&0&0&\cdots\\
\end{bmatrix}
\end{align}
\end{widetext}
where the ``$\cdots$" indicate that the last eight columns are repeated, replacing 
$c^\dagger_{1\uparrow}\rightarrow c^\dagger_{1\downarrow}$ and $c_{1\uparrow}\rightarrow c_{1\downarrow}$.

The second MPO matrix is:
\begin{widetext}
\setcounter{MaxMatrixCols}{30}
\begin{align}
\hat{M}_2 = \begin{bmatrix}
I_2 & V_{pp} (n_2)^2  & 0    & 0   & 0&0&0&n_2&0&0&0&c^\dagger_{2\uparrow}&0&0&0&c_{2\uparrow}&\cdots\\
0   & I_2             & 0    & 0   & 0   &0&0&0&0&0&0&0&0&0&0&0&\cdots\\
0   & \Omega^{22}_{11} n_2 & I_2  & 0   & 0   &0&0&0&\Omega^{23}_{11} c^\dagger_{2\uparrow} & 0&0&0&-\Omega^{32}_{11} c_{2\uparrow} &0&0&0&\cdots\\
0   & \Omega^{22}_{21} n_2 & 0    & I_2 & 0   &0&0&0&\Omega^{23}_{21} c^\dagger_{2\uparrow} & 0 &0&0&-\Omega^{32}_{21}c_{2\uparrow}&0&0&0&\cdots\\
0   & 0               & 0    & 0   & I_2 &0 &0 &0 &0 &0 &0 &0 &0 &0 &0 &0&\cdots \\
0   & 0               & 0    & 0   & 0   &0 &0 &0 &0 &0 &0 &0 &0 &0 &0 &0&\cdots \\
0   & 0               & 0    & 0   & 0   &0 &0 &0 &0 &0 &0 &0 &0 &0 &0 &0&\cdots \\
0   & 0               & 0    & 0   & 0   &0 &0 &0 &0 &0 &0 &0 &0 &0 &0 &0&\cdots \\
0   & c_{2\uparrow}   & 0    & 0   & 0   &0 &0 &0 &0 &0 &0 &0 &0 &0 &0 &0&\cdots \\
0   & 0               & 0    & 0   & 0   &0 &0 &0 &F_2 & 0 &0 &0 &0 &0 &0 &0&\cdots \\
0   & 0               & 0    & 0   & 0   & c_{2\uparrow}   &0 &0 &0 &0 &F_2 &0 &0 &0 &0 &0&\cdots \\
0   & 0 & 0    & 0   & 0   &0 & 0 &0 &0 &0 &0 &0 &0 &0 &0 &0&\cdots \\
0   & c^\dagger_{2\uparrow} & 0    & 0   & 0   &0 &0 &0 &0 &0 &0 &0 &0 &0 &0 &0&\cdots \\
0   & 0 & 0    & 0   & 0   &0 &0 &0 &0 &0 &0 &0 &F_2 &0&0 &0&\cdots \\
0   & 0 & 0    & 0   & 0   &0 & c^\dagger_{2\uparrow} &0 &0 &0 &0 &0 &0 &0 &F_2 &0&\cdots \\
0   & 0 & 0    & 0   & 0   &0 & 0 &0 &0 &0 &0 &0 &0 &0 &0 &0&\cdots \\
\vdots & \vdots & \vdots & \vdots & \vdots & \vdots & \vdots & \vdots & \vdots & \vdots & \vdots & \vdots & \vdots & \vdots & \vdots & \vdots
\end{bmatrix}
\end{align}
\end{widetext}
where $F_2=(-1)^{n_2}$ is a fermion string operator. We include this detail to note that, at least in our
ITensor implementation, the operators we denote here as $c$ and $c^\dagger$ only anticommute when acting
on the same site, so the additional $F$ operators must be included between $c^\dagger$ and $c$ pairs acting
on different sites. If the anticommutation bookkeeping is done in a more automatic way, where $c$ and $c^\dagger$
really do anticommute across different sites, one would replace these
$F$ operators with identity operators.

The third, and last MPO matrix on this $N_o=3$ slice is:
\setcounter{MaxMatrixCols}{30}
\begin{align}
\hat{M}_3 = \begin{bmatrix}
I_3 & V_{pp} (n_3)^2  & U^{33}_{p1} n_3 & U^{33}_{p2} n_3  \\
0   & I_3             & 0    & 0 \\
0   & \Omega^{33}_{11} n_3 & X_{11}  & X_{12}  \\
0   & \Omega^{33}_{21} n_3 & X_{21}    & X_{22} \\
0   & 0               & U^{11}_{p1} I_3    & U^{11}_{p2} I_3  \\
0   & 0               & U^{21}_{p1} I_3    & U^{21}_{p2} I_3  \\
0   & 0               & U^{12}_{p1} I_3    & U^{12}_{p2} I_3  \\
0   & 0               & U^{22}_{p1} I_3    & U^{21}_{p2} I_3  \\
0   & c_{3\uparrow}   & 0    & 0  \\
0   & 0               & 0    & 0  \\
0   & 0               & U^{13}_{p1} c_{3\uparrow} & U^{13}_{p2} c_{3\uparrow}  \\
0   & 0               & U^{23}_{p1} c_{3\uparrow} & U^{23}_{p2} c_{3\uparrow}  \\
0   & c^\dagger_{3\uparrow}   & 0    & 0  \\
0   & 0                       & 0    & 0  \\
0   & 0               & U^{31}_{p1} c^\dagger_{3\uparrow} & U^{31}_{p2} c^\dagger_{3\uparrow}  \\
0   & 0               & U^{32}_{p1} c^\dagger_{3\uparrow} & U^{32}_{p2} c^\dagger_{3\uparrow}  \\
\vdots & \vdots & \vdots & \vdots
\end{bmatrix}
\end{align}

To make a well-defined MPO for a finite system, the first and last MPO tensors are contracted 
with the a boundary vector to the left of the first site:
\begin{align}
\vec{L}^{T} = 
\begin{bmatrix}
1 &0&0&0
\end{bmatrix}
\end{align}
and a boundary vector to the right of the last site:
\begin{align}
\vec{R} = 
\begin{bmatrix}
\,0\  \\
\,1\  \\
\,0\  \\
\,0\  \\
\,\vdots\ 
\end{bmatrix} \ .
\end{align}

To explain the design of the MPO above in words (recalling that it is a concrete example for the 
case $N_o=3$ and $D=2$, so that the row and column numbers are specific to that case):
\begin{enumerate}
\item Row 1 of each MPO matrix holds operators which begin an operator ``string'' on that site (these are the ``starting'' operators in a finite-state automaton picture of an MPO, Ref.~\onlinecite{Crosswhite:2008}).
\item The identity operator at element (2,2) of each matrix trails a completed string of operators (the ``done'' state in an automaton picture).
\item Rows and columns 3 and 4 correspond to the $\alpha$ indices formed from the SVDs in the compression algorithm (for general $D$ this would be rows and columns $3,4,\ldots,D+2$). For sites 1 and 2, operators from previous slices either connect with elements of $\Omega$ to form a completed operator string or are passed through to the next site. On site 3 (more generally site $N_o$), the $X$ matrix appears, transforming incomplete operator strings from the $\alpha_{p-1}$ basis into the $\alpha_{p}$ basis.
\item Columns 5--8 of $\hat{M}_1$, and rows and columns 5--8 of $\hat{M}_2$ collect pairs of operators on sites 1 and 2. In rows 5--8 of $\hat{M}_3$, these operator pairs get multiplied by elements of 
$U$ on site 3 to begin a new operator string connecting to a different slice.
\item Rows and columns 9 and 10 of $\hat{M}_1$ and $\hat{M}_2$ multiply nearly-complete operator strings from a 
previous slice by an element of $\Omega$ and
a $c^\dagger_{\uparrow}$ operator. However these operator strings must be carried on to one of the remaining sites in the slice (sites 2 or 3) to be matched with the $c_{\uparrow}$ operators in column 2 of $\hat{M}_2$ and $\hat{M}_3$.
\item Rows and columns 11 and 12 begin operator strings starting with a $c^\dagger_\uparrow$ on either site 1 or 2, which will be paired with an element of $U$ and a $c_{3\uparrow}$ operator on site 3 to begin a new operator string.
\end{enumerate}
The pattern of columns 9--12 of $\hat{M}_1$ and $\hat{M}_2$ repeats three more times, replacing $c^\dagger_\uparrow$ with $c_\uparrow$, $c^\dagger_\downarrow$, and $c_\downarrow$.

Note that in the third MPO matrix (more generally, the matrix on site number $N_o$ within a slice) we weighted new
operator strings with elements of $U^{il}$ instead of elements of $X$ as in the $N_o=1$ MPO Eq.~(\ref{eqn:MPO1_gen}). 
This was for convenience as the elements $U^{(p)\,il}_{p\alpha_p}$ for \mbox{$i,l=1,2,\ldots,N_o$} and \mbox{$\alpha_p=1,2,\ldots,D$} 
correspond to the last $N_o^2$ rows of $X^{(p)}$, and listing these rows of $X^{(p)}$ would be unwieldy in the current
notation.

\end{document}